\documentclass{aastex63}
\usepackage{graphicx}
\usepackage{CJK}

\begin{document}
\begin{CJK*}{UTF8}{gbsn}

\title{The metallicity distribution in the LMC and the SMC based on the Tip-RGB colors}

\correspondingauthor{Biwei Jiang}
\email{bjiang@bnu.edu.cn}

\author[0009-0001-5020-4269]{Ying Li (李颖)}
\affiliation{Institude for Frontiers in Astronomy and Astrophysics, Beijing Normal University, Beijing 102206, China}
\affiliation{Department of Astronomy, Beijing Normal University, Beijing 100875, China}

\author[0000-0003-3168-2617]{Biwei Jiang (姜碧沩)}
\affiliation{Institude for Frontiers in Astronomy and Astrophysics, Beijing Normal University, Beijing 102206, China}
\affiliation{Department of Astronomy, Beijing Normal University, Beijing 100875, China}

\author[0000-0003-1218-8699]{Yi Ren (任逸)}
\affiliation{College of Physics and Electronic Engineering, Qilu Normal University, Jinan 250200, China}

\begin{abstract}
The color index $(J-K)_0$ of tip-red giant branch (TRGB) is used to study the metallicity distribution in the Large and Small Magellanic Cloud. With the most complete and pure sample of red member stars so far, the areas are divided into 154 and 70 bins for the LMC and SMC respectively with similar number of stars by the Voronoi binning. For each bin, the position of TRGB on the near-infrared color-magnitude diagram, specifically $(J-K)_0/K_0$, is determined by the Poison-Noise weighted method. Converting the color index of TRGB into metallicity, the metallicity gradients in the LMC and the SMC are obtained in four major directions. For the LMC, the gradient to the north is $-0.006 \pm 0.004$ dex kpc$^{-1}$, to the south  $-0.010 \pm 0.005$ dex kpc$^{-1}$, to the east $-0.006 \pm 0.003$ dex kpc$^{-1}$, and to the west  $-0.010 \pm 0.003$ dex kpc$^{-1}$. The farthest distance extends to 16 kpc. For the SMC, the gradients to the north, south, east, and west are $-0.017 \pm 0.031$ dex kpc$^{-1}$, $-0.016 \pm 0.007$ dex kpc$^{-1}$, $-0.003 \pm 0.002$ dex kpc$^{-1}$, and $-0.004 \pm 0.003$ dex kpc$^{-1}$, respectively. The farthest distance for the SMC extends to 27 kpc. The gradient is large from the center to 1 kpc.
\end{abstract}

\keywords{\href{http://astrothesaurus.org/uat/1371}{Red giant tip(1371)}; \href{http://astrothesaurus.org/uat/990}{Magellanic Clouds(990)}; \href{http://astrothesaurus.org/uat/1031}{Metallicity(1031)}
}

\section{Introduction} \label{sec:intro}

The Large and Small Magellanic Clouds, i.e. LMC and SMC, are the largest and nearest Milky Way (MW) dwarf satellite galaxies, situated at a distance of $\sim$50 and 61 kpc respectively (\citealt{2014AJ....147..122D}; \citealt{2015AJ....149..179D}), which are separated by around $\sim$20 kpc \citep{2018ApJ...864...55Z}. The Magellanic Clouds are a pair of interacting galaxies whose mutual interaction is thought to be the reason for their metallicity gradients \citep{2009A&A...506.1137C}. It should be noted that the metallicity gradient in the outer disk may due to the age distribution of the stars \citep{2011AJ....142...61C}. 
\citet{2008AJ....135..836C} found that the age-metallicity relationship observed in their studied disk field of the LMC aligns well with a single global trend of the disk, while the outermost field is more metal-poor on average because it contains a lower fraction of relatively young stars (age $<$ 5 Gyr).

In a recent study, \cite{2021MNRAS.507.4752C} identified a mildly metal-rich bar in the LMC compared to its outer disc, suggesting the presence of a subtle gradient in metallicity (-0.008 $\pm$ 0.001 dex kpc$^\mathrm{-1}$) from the galaxy's core to a radius of 6 kpc. The mean metallicity they found is [Fe/H] = -0.42.
Similarly, \cite{2009A&A...506.1137C} utilized asymptotic giant branch (AGB) stars in the LMC to discern a metallicity gradient of -0.047 $\pm$ 0.003 dex kpc$^\mathrm{-1}$, extending to a radius of $\sim$8 kpc from the center.
Analyzing the red giants of the LMC, \cite{2016MNRAS.455.1855C} estimated a gradient of -0.049 $\pm$ 0.002 dex kpc$^\mathrm{-1}$ spanning the LMC disk up to $\sim$4 kpc.
\cite{2021ApJ...909..150G} characterized the LMC as an inclined thin disk using the Gaia/DR2 measurements. Their findings indicate a gradual metallicity gradient of -0.048 $\pm$ 0.001 dex kpc$^\mathrm{-1}$ up to $\sim$12${\degr}$ from the dwarf's center. Notably, they observed pronounced asymmetries and plateau features beyond the central region and the intricate metallicity structure within the Magellanic Bridge region. Additionally, \cite{2008AJ....135..836C} used the CaII triplet spectroscopy to deduce metallicity patterns. Their analysis revealed that the mean metallicity remains relatively constant ([Fe/H] $\sim$ -0.5 dex) until 6${\degr}$, whereas [Fe/H] $\sim$ -0.8 dex at 8${\degr}$ is lower compared to other areas of the disk. \citet{2016AcA....66..269S} obtain the integral gradient of the LMC of $-0.019 \pm 0.002$ dex kpc$^{-1}$, and they provide a segmented gradient with $-0.029 \pm 0.002$ dex kpc$^{-1}$ in the inner 5 kpc and $-0.030 \pm 0.003$ dex kpc$^{-1}$ beyond 8 kpc, and in their work, there is no gradient in-between. \citet{2021ApJS..252...23S} use the red giants from APOGEE survey to yield the metallicity gradient of $-0.026 \pm 0.002$ dex kpc$^{-1}$. Using SDSS-IV/APOGEE-2S data, \citet{2023arXiv230912503P} obtain the gradient of the LMC being $-0.0380 \pm 0.0022$ dex kpc$^{-1}$. By exploring the outer disk of the LMC, \citet{2011AJ....142...61C} argue that there exists a metallicity gradient in the most metal-rich populations outwards of $\sim$7kpc.

\cite{2008AJ....136.1039C} identified a negative metallicity gradient in the SMC through spectroscopic measurements.
Building upon this, \cite{2020MNRAS.497.3746C} crafted high-spatial-resolution metallicity map spanning approximately $\sim$42 deg$^\mathrm{2}$ across the expanse of the SMC. Their investigation revealed a gentle metallicity gradient (-0.031 $\pm$ 0.005 dex deg$^\mathrm{-1}$) extending from the galactic center to the radius of 2${\degr}$ to 2.5${\degr}$. Beyond this range, a phase of constant metallicity follows, spanning around $\sim$ 3.5${\degr}$ to 4${\degr}$. Notably, their findings demonstrate that the metallicity gradient in the SMC exhibits a radial asymmetry. The metallicity gradient of the SMC derived by \citet{2023arXiv231014299P} is $-$0.0546 $\pm$ 0.0043 dex kpc$^{-1}$ by using the high resolution H-band APOGEE-2S spectra of RGB stars.

The tip of the red giant branch (TRGB) is the sharp cut-off on the color-magnitude diagram occurring at the bright end of the red giant branch.
The magnitude of the TRGB in the $I$ band has been employed as a standard candle (\citealt{1995AJ....109.1645M}; \citealt{2004ApJ...606..869B}).
The concept of using the TRGB for the distance measurement was first proposed by \cite{1944ApJ...100..137B}.
\cite{2017RNAAS...1...31L} contributed to the understanding of  the TRGB magnitudes in the $J$ band regarding of stellar age and metallicity.

The potential of using infrared photometry of TRGB to study galactic structure is proved in previous studies. \cite{1990ApJ...362..503B} use the Galactic and M31 cluster spectra to derive the linear relations between color index and metallicity with new calibration of infrared colors ($V-K$, $J-K$, and the CO spectral index) versus [Fe/H], and provide an estimate of [Fe/H] accurate to about 15\%.
\cite{2000AJ....119..727B} point that the advantage of using $J-K$ as a metallicity indicator is its insensitivity to interstellar reddening.
Meanwhile, \cite{2004A&A...424..199B} derived fresh empirical calibrations, offering robust connections between the TRGB magnitude in the $I$, $J$, $H$, and $K$ bands and metallicity.

In this work, we try to analyze the metallicity distribution of the Magellanic Clouds with the near-infrared brightness and color of the member stars' TRGBs. The star catalog is taken from the recent sample of \citet{2021ApJ...923..232R}, which has removed the foreground dwarf stars by proper motion, parallax and their obvious branch in the near-infrared $J-H/H-K$ diagram with the 2MASS photometry and the Gaia/EDR3 measurements. In addition, the interstellar extinction is calculated and corrected. The combination of astrometric and astrophysical measurements brings about the most pure and complete sample of the red stars in the MCs.

This paper is organized as follows. Section 2 introduces the data, followed by Section 3 to describe the method of detecting the TRGB. The results and discussions are presented in Section 4, and the summary in Section 5.

\section{Data} \label{sec:data}

\begin{figure}
	\centering
    \includegraphics[scale=0.9]{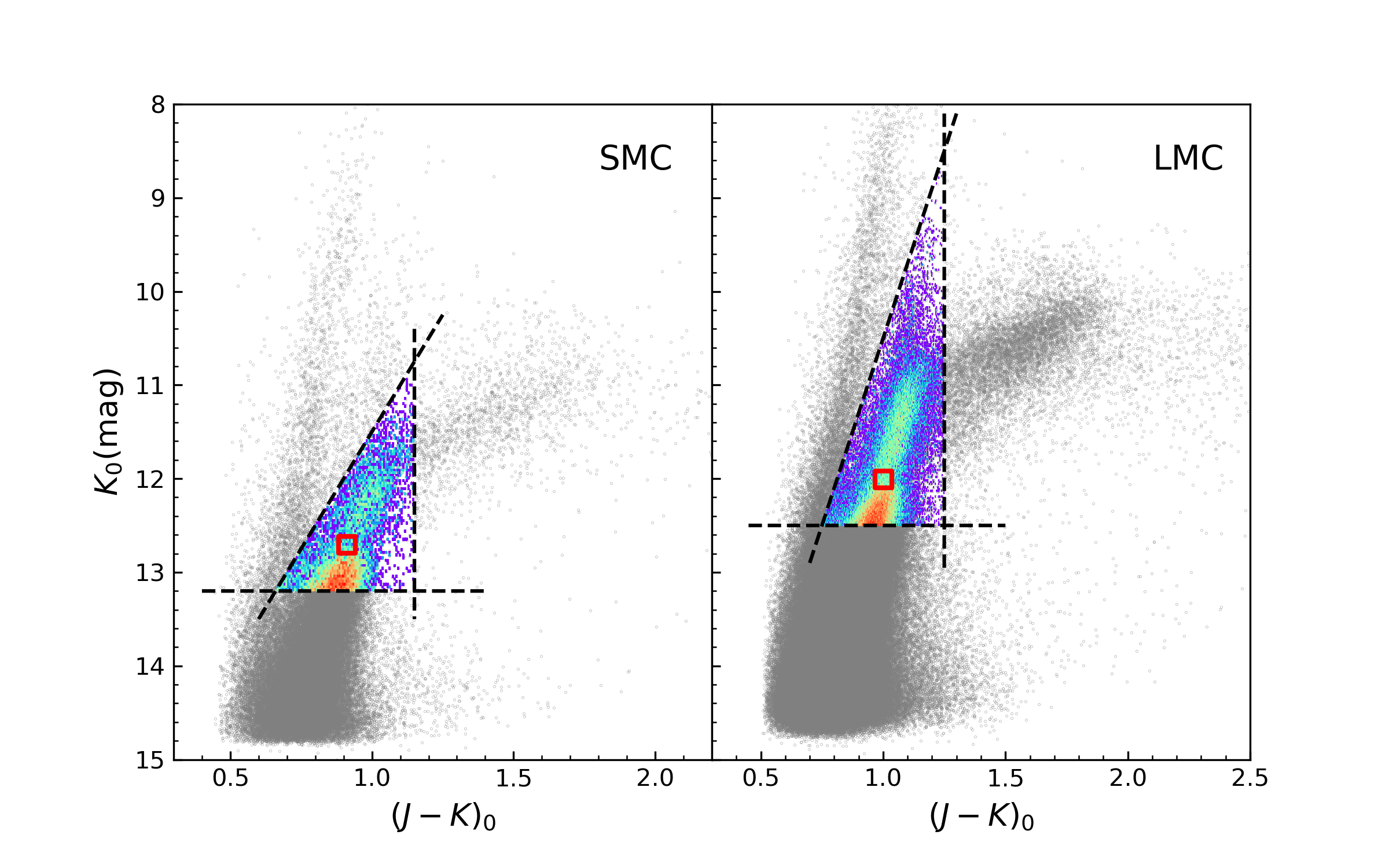}
	\caption{The $(J-K)_{0}$ vs. $K_{0}$ diagram of the SMC and the LMC member stars. The dashed triangle indicates the area used to determine the color and brightness of TRGB, and the red box denotes the position of the TRGB in either galaxy. \label{fig:MCs_delect}}
\end{figure}

Our study relies on near-infrared photometry of the stars in the Magellanic Clouds (MCs) area. The sample is taken from the catalogs constructed by \citet{2021ApJ...923..232R}. In their work, \citet{2021ApJ...923..232R} removed the foreground dwarf stars first by the  measurements of proper motions and parallax from Gaia early data release 3 (EDR3), which aligns with the approach of Yang et al.(\citeyear{2019A&A...629A..91Y}, \citeyear{2021A&A...646A.141Y}, \citeyear{2020A&A...639A.116Y}). Specifically, the stars whose proper motions are beyond the range of the MCs are rejected, and those with the distance calculated from the parallax are also removed if less than 35 kpc. In addition, the near-infrared color-color diagram is applied to further remove the foreground dwarf stars by their locations obviously different from red giant stars \citep{2021ApJ...907...18R}. 
The $JHK$ magnitudes used in their work are retrieved from the 2MASS PSC  \citep{2006AJ....131.1163S} in which the photometry quality is required to be “AAA”, i.e. the best quality in all the three bands.
The 2MASS project used two 1.3 m telescopes located at Mt. Hopkins and CTIO, resulting in photometric systematic uncertainties of better than 0.03 mag \citep{2022MNRAS.511.1317C}.
\cite{2021ApJ...923..232R} converted the 2MASS magnitudes into the UKIRT system using the transformation equations derived by \cite{2009MNRAS.394..675H}. In their catalog, there are 402,300 member stars identified in the LMC and 66,573 in the SMC. Figure \ref{fig:MCs_delect} displays the $(J-K)_{0}$ versus $K_{0}$ diagram of the member stars from \cite{2021ApJ...923..232R}. With the high accuracy of Gaia astrometric measurements and the novel method to remove the foreground stars, this sample is both complete and pure for the bright red giants, AGB stars, and red supergiants in the MCs.

Since this work focuses on the Tip-RGB,  the sample of \citet{2021ApJ...923..232R} is reduced to include only the stars in close proximity to the Tip-RGB in the color-magnitude diagram to exclude the unnecessary interference of other objects. In Figure \ref{fig:MCs_delect}, the regions of the stars selected for later analysis are marked by the dashed lines.
This selection leads to 49,695 member stars in the LMC (hereafter referred to as SampleL), and 9,285 member stars in the SMC (hereafter referred to as SampleS) kept for detecting the TRGB.

\section{Method} \label{sec:method}

\subsection{The Voronoi Binning} \label{sec:Voronoi Binning}

Considering the very uneven distribution of stars in the field, the usual bin with an equal space interval is inappropriate. Instead, we use the Voronoi tessellation to divide the entire sky area into multiple regions each of which contains approximately the same number of stars. The Voronoi tessellation is a method of spatial segmentation that has applications in many fields of science and technology and social sciences. It has many applications in astronomy, such as spectrography data analysis, image processing, adaptive smoothing, segmentation, signal-to-noise ratio balancing, and so on \citep{2021inas.book...57V}.

\begin{figure}
    \centering
    \includegraphics[scale=0.8]{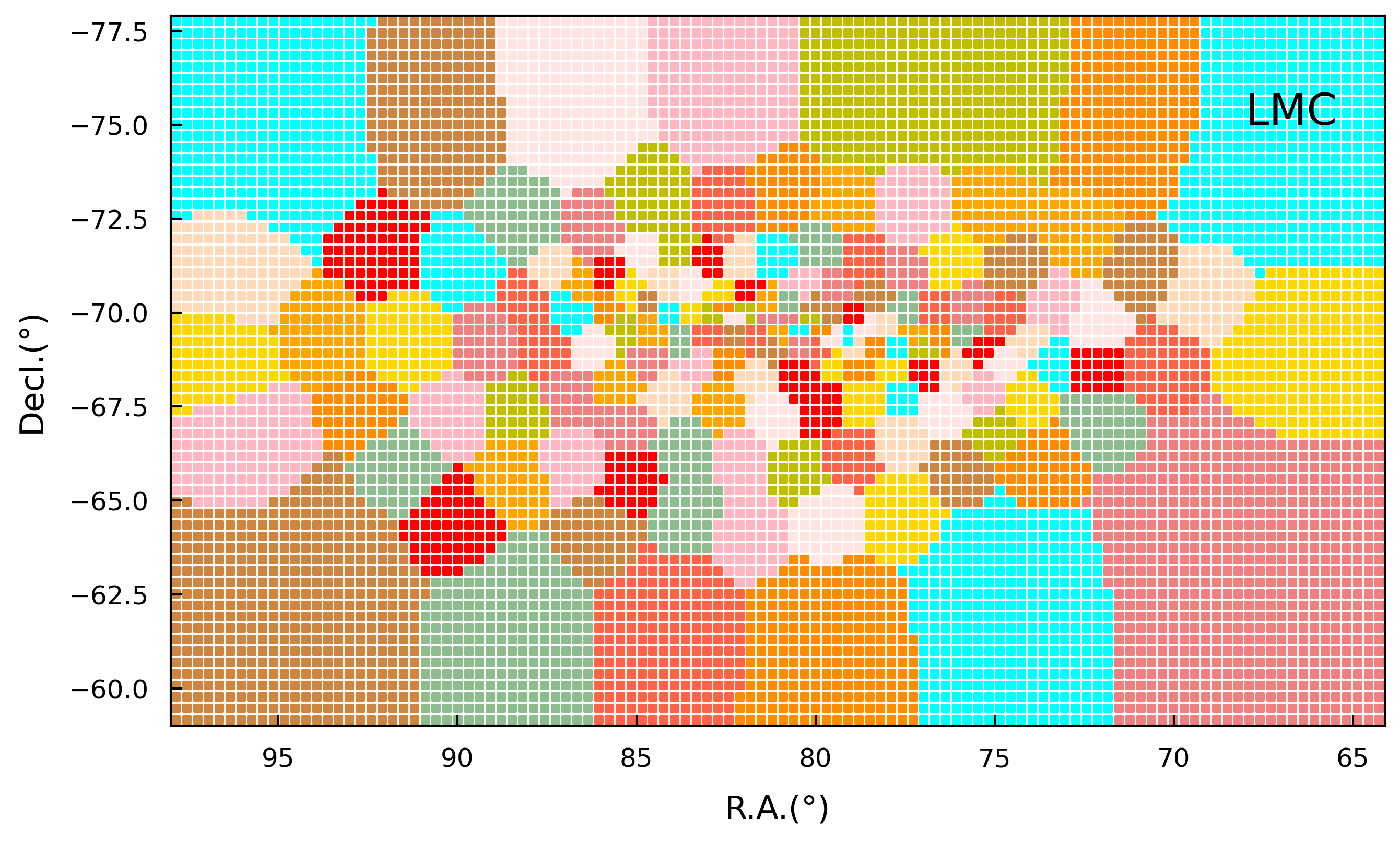}
    \caption{The Voronoi bins for the LMC. The smallest squares represent the initial bin unit and the irregular polygon including multiple bin units of the same color denotes one final bin with about 300 stars by the Voronoi binning method. In total, there are 154 bins.}
    \label{fig: LMC-voronoi}
\end{figure}

\begin{figure}
    \centering
    \includegraphics[scale=0.5]{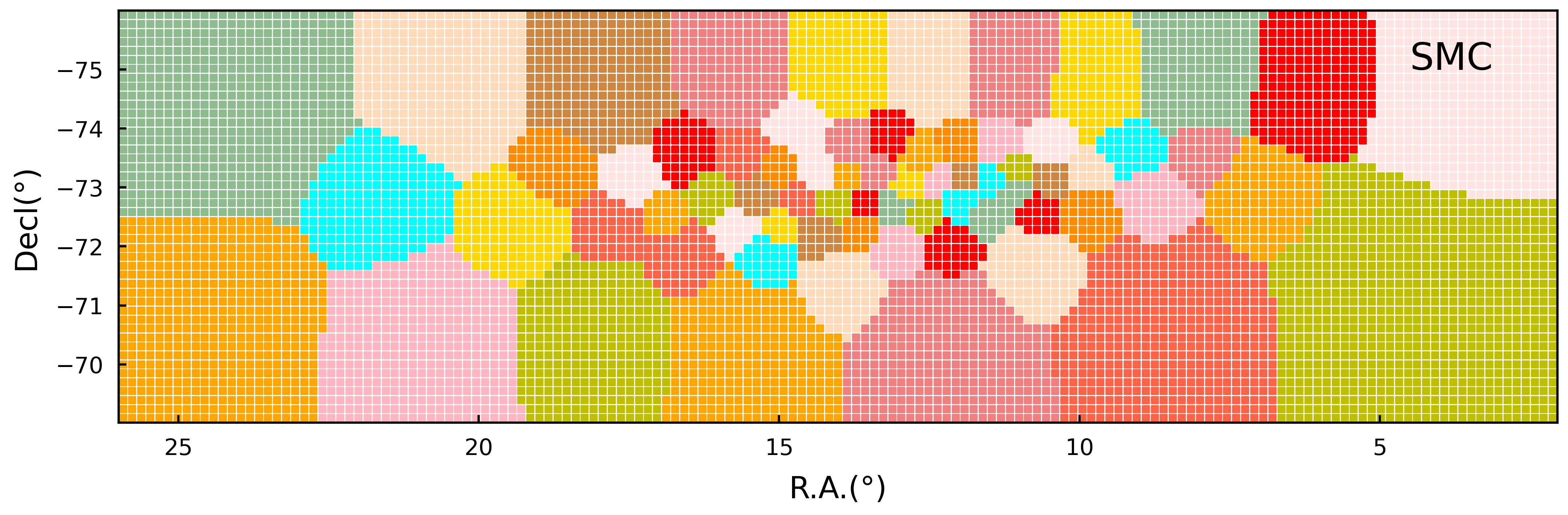}
    \caption{The same as Figure \ref{fig: LMC-voronoi}, but for the SMC and the number of stars in one bin is about 120. In total, there are 70 bins.} 
    \label{fig: SMC-voronoi}
\end{figure}

In the first step, the entire sky area of study is divided into many "pixels", and each "pixel" has an equal size but may contain a different number of stars. Specifically,
the LMC is bounded by $64.112\degr < R.A. < 98.000\degr$, $-77.912\degr < Decl. < -59.014\degr$, and  divided into 112$*$62 "pixels" with each pixel of 0.3$\degr$ by 0.3$\degr$. Similarly,
the SMC is bounded by $2.043\degr < R.A. < 25.994\degr$, $-75.997\degr < Decl. < -69.015\degr$, and divided into 159$*$46 "pixels" with each pixel of 0.15$\degr$ by 0.15$\degr$.
Then an expected number of stars, 300 for the LMC and 120 for the SMC here, is set for one bin. The expected bins are produced by the adaptive Voronoi binning technique proposed by \cite{2006MNRAS.368..497D}, which is a generalization of the \cite{2003MNRAS.342..345C} Voronoi binning algorithm. For the objective to have approximately 300 stars within one bin, the algorithm commences with a "pixel" and progressively incorporates neighboring pixels until the bin accommodates roughly 300 stars. This process starts from the nearest unbinned pixel, provided that
(a) no empty or overlapping bins,
(b) bins meet predefined roundness standards,
and (c) minimal spread of star counts in each bin.
Consequently, the LMC is divided into 154 bins, denoted as binNum0 to binNum153, each containing approximately 300 stars. Similarly, the SMC is divided into 70 bins, marked as binNum0 to binNum69, with each bin encompassing roughly 120 stars.
Figure \ref{fig: LMC-voronoi} and Figure \ref{fig: SMC-voronoi} present the outcomes for the LMC and the SMC respectively.

\subsection{Detecting the TRGB by the PN method}

In the $(J-K)_0$ vs. $K_0$ diagram, the star density exhibits a gradual decrease and subsequent increase from the RGB to the AGB, with the TRGB positioned at the point of lowest density. Some sophisticated techniques are developed to measure the brightness and colors of the TRGB.
\citet{2018AJ....156..278G} employed a simple edge detection technique based on the comparison of the star count difference in two adjacent bins weighted by the estimated Poisson noise.
\citet{2021ApJ...907...18R} calculated the saddle points to determine the brightness and color of TRGB in both M31 and M33. Another method making use of the Sobel filter is proven effective in measuring the TRGB brightness \citep{1993ApJ...417..553L}. However, in certain fields of the Magellanic Clouds, measuring TRGB brightness becomes challenging due to the Sobel filter's tendency to produce multiple peaks in close proximity to the expected luminosity function cutoff.
\cite{2018AJ....156..278G} used the Maximum Likelihood Algorithm (MLA)  \citep{2002AJ....124..213M} in the LMC and the SMC, but for many fields, the TRGB brightness differs by more than 1 mag from the expected value.

We adopt the Poisson Noise weighted star counts difference in two adjacent bins suggested by \citet{2018AJ....156..278G}, hereafter referred to as the PN method, to identify the TRGB in both the LMC and the SMC. The PN method offers a distinct advantage, especially in cases where galaxies lack prominent AGBs compared to the saddle point method.
The Gaussian kernel density estimation is used to calculate the number density distribution in the $(J-K)_0$  vs. $K_0$ diagram. In the 3D density distribution, the $K_{0}$-axis is sliced with a step of 0.01. The highest point in each slice is taken as the maximum value, all of which consist the ridge of 3D density distribution.
The ridge is then projected along the direction parallel to $(J-K)_{0}$ and to $K_{0}$ to obtain the probability density function (PDF) of $(J-K)_{0}$ and $K_{0}$, respectively.
\cite{2002AJ....124.3222B} point out that TRGB has a certain width in brightness so that the position of TRGB detected by the PN method would be fainter than the true TRGB. We thus
consider the PN-detected $K_{\rm 0}$ magnitude to be the faint end of TRGB, while the bright end of TRGB is taken to be the first brighter local minimum of the PN filter, and the average of the bright and faint ends is the true magnitude of TRGB.

\begin{figure}
	\centering
    \includegraphics[scale=0.4]{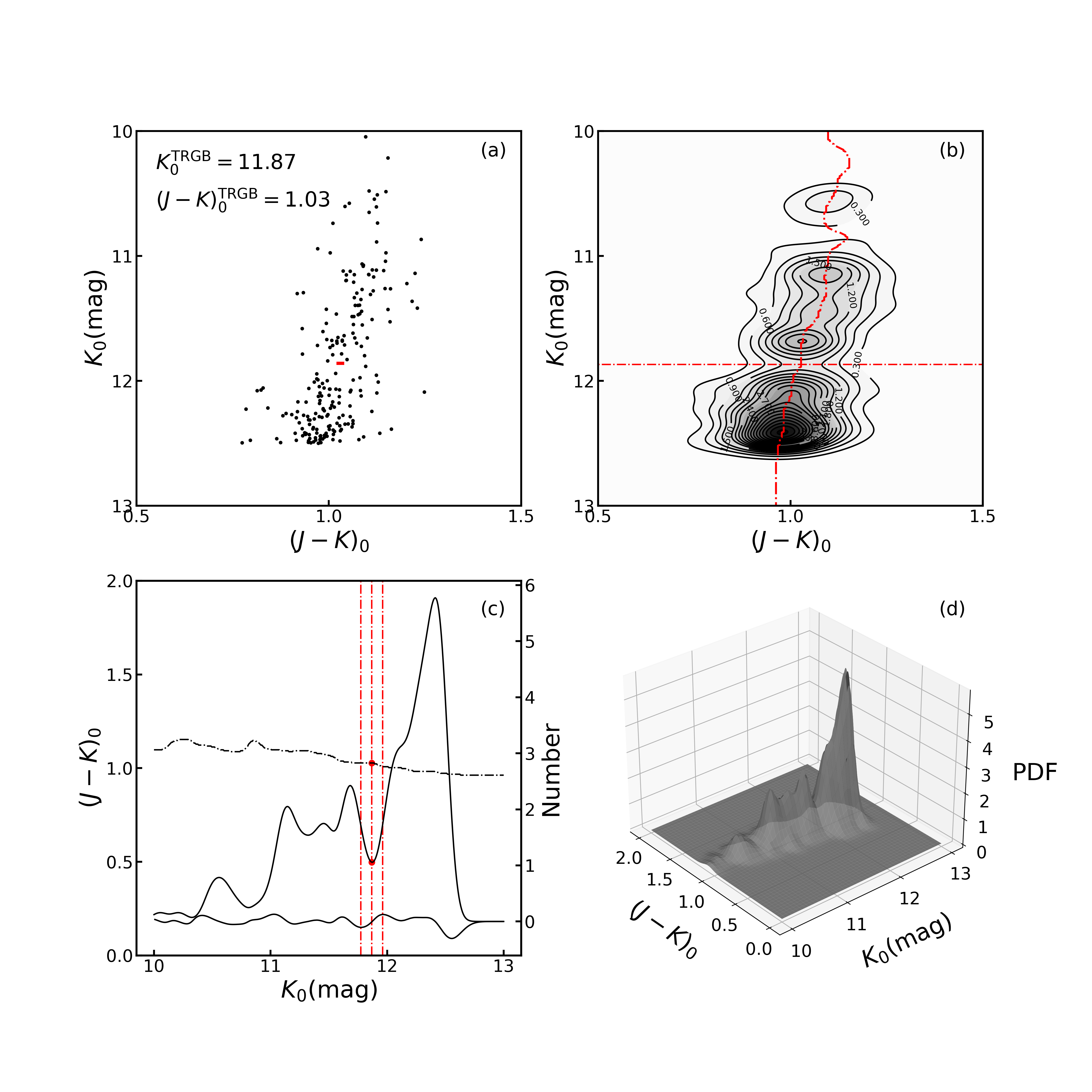}
	\caption{The example to show the PN method to determine the TRGB position by the binNum4 in the LMC. (a) the color-magnitude diagram of the member stars. The red box indicates the position of the TRGB with the error bars fitted in Figure \ref{fig:bootstrap2000-example} and the values are displayed in the upper left corner; (b) $PDF$ of the member stars, where the red line shows the $K_{0}$  vs. $(J-K)_{0} $ relation of the ridge line; (c) The black dashed-dotted line shows the $PDF_{\rm ridge}$ vs. $(J-K)_{0}$ relation, the black solid line shows the $PDF_{\rm ridge}$ vs. $K_{0}$ relation. The red dashed-dotted lines on the right and left mark the bright and faint ends of TRGB, respectively, and the middle one for the true position of TRGB; (d) the 3D density distribution of member stars.  \label{fig:PN_example}}
\end{figure}

The case of binNum4 in the LMC is displayed in Figure \ref{fig:PN_example} as an example of how the Poisson Noise (PN) method is employed to determine the TRGB location:
(a)The color-magnitude diagram of the member stars where the red box indicates the position of the TRGB;
(b)$PDF$ of the member stars, where the red dashed line shows the $K_{0}$  vs. $(J-K)_{0} $ relation of the ridge line;
(c)The black dashed-dotted line shows the $PDF_{\rm ridge}$ vs. $(J-K)_{0}$ relation, and the black solid line shows the $PDF_{\rm ridge}$ vs. $K_{0}$ relation. The red dashed-dotted lines on the right and left indicate the bright and faint ends of TRGB, respectively, and the middle one indicates the position of TRGB;
(d)The 3D density distribution of member stars.
The derived position of TRGB is $K_{0} = 11.87, (J-K)_{0}=1.03$. This represents the outcome of a single calculation.

\subsection{The Bootstrap error analysis}

The uncertainty of the TRGB position in the CMD is estimated by the bootstrap method. The bootstrap resampling method, introduced by \citet{Efron1979}, generates a large number of data sets, each with the same number of data points randomly drawn from the original sample. Each drawing is made from the entire data set and some points may be duplicated while some may be missing. We apply the no-relaxation sampling method (parameter: replace=0), taking 80\% of the total number to which the PN method is applied to calculate the color index and magnitude of TRGB. This process is repeated 2000 times, yielding 2000 sets of TRGB color index and magnitude values. From this collection, the expected value $\mu_{(J-K)_{0}}$ and standard deviation $\sigma_{(J-K)_{0}}$ are determined by a Gaussian fitting to the distribution of $(J-K)_{0}$. Afterwards, the results falling within 1 standard deviation $\sigma_{(J-K)_{0}}$ are selected to fit $K_{0}$, thus yielding the expected value $\mu_{K_{0}}$ and its standard deviation $\sigma_{K_{0}}$. This approach ensures robustness and reliability in estimating the key parameters of interest.

\begin{figure}
    \centering
    \includegraphics[scale=0.5]{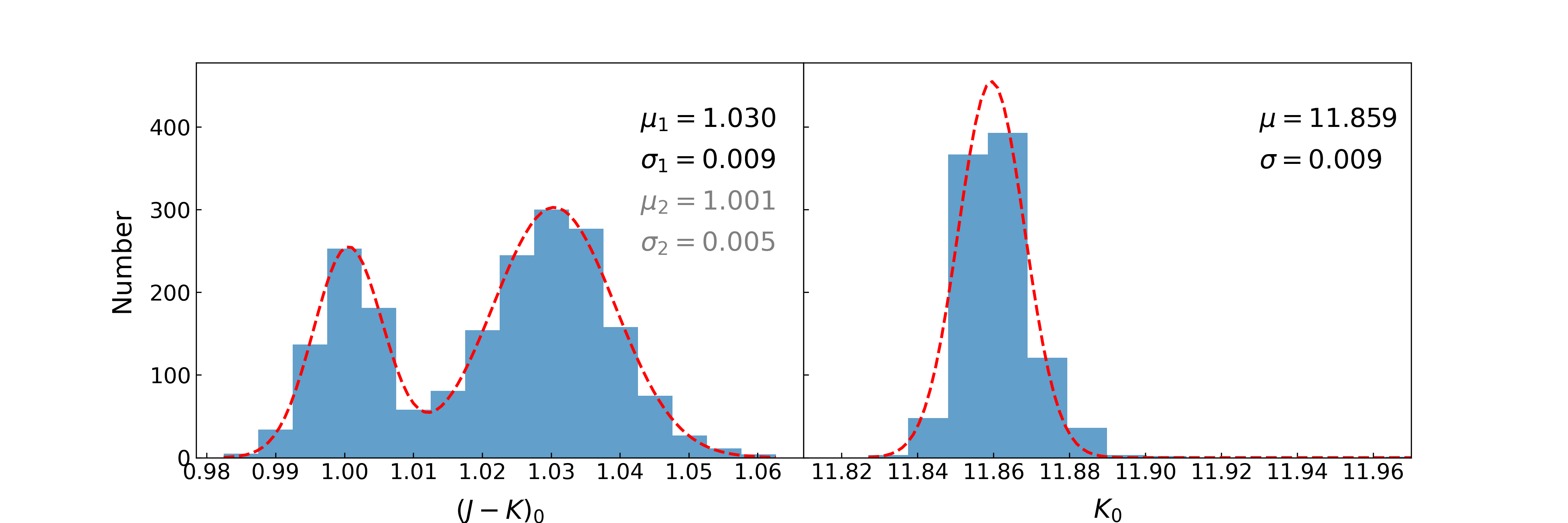}
    \caption{Distribution of the 2000 Bootstrap re-sampling results in $(J-K)_{0}$ and $K_{0}$ for the bin in Figure \ref{fig:PN_example}, and the results from Gaussian fitting are displayed.}
    \label{fig:bootstrap2000-example}
\end{figure}

The case of binNum4 of the LMC mentioned in the previous section is again taken as an example. In Figure \ref{fig:bootstrap2000-example}, the left panel displays the fitting result of $(J-K)_{0}$ over 2000 iterations. If the distribution resembles a single Gaussian,  $\mu_{(J-K)_{0}}$ and $\sigma_{(J-K)_{0}}$ are directly retracted. In case of double or multiple Gaussian distributions, the counts within $3\sigma$ of each Gaussian are compared and the one with the highest count is selected for the mean $\mu$ and its error $\sigma$. The bold font represents the selected $\mu$ and $\sigma$. The right panel of Figure \ref{fig:bootstrap2000-example} demonstrates the results within $\mu_{(J-K)_{0}} \pm \sigma_{(J-K)_{0}}$ for fitting $K_{0}$ and generating the distribution of $K_{0}$. The determination of $\mu_{K_{0}}$ and $\sigma_{K_{0}}$ followed the same procedure as $(J-K)_{0}$. Therefore, for binNum4 in the LMC, the TRGB position is determined to be: $(J-K)_{0} = 1.030 \pm 0.009$, $K_{0} = 11.859 \pm 0.009$. It deserves to mention that the appearance of multiple Gaussian distributions may come from the insignificant feature of TRGB in the bin.

\section{Result and Discussion} \label{sec:results}

\subsection{TRGB of the LMC and the SMC} \label{4.1TRGB in L and S}

With all the member stars from SampleL and SampleS of \cite{2021ApJ...923..232R} and the methods described above, the position of TRGB as an entity in the galaxy is yielded as follows: $K_{0} = 12.007 \pm 0.006$, $(J-K)_{0} = 1.002 \pm 0.001$ for the LMC, and $K_{0} = 12.707 \pm 0.006$ and $(J-K)_{0} = 0.912 \pm 0.002$ for the SMC. In comparison, \cite{2021ApJ...923..232R} derived the TRGB position as $K_{0}=12.00$ and $(J-K)_{0}=1.00$ for the LMC, and $K_{0}=12.71$, $(J-K)_{0}=0.91$ for the SMC by using the same catalog. The highly consistent results confirm the correctness of our method and the difference implies the error being smaller than 0.01.

Previously, \cite{2015AJ....149..117M} used the CPAPIR camera at the CTIO 1.5-m telescope to obtain a near-infrared synoptic survey of the central region of the LMC from which they derived the TRGB position of $K_{\rm S} =12.11\pm 0.01$ and $(J-K)=1.12$.
In a similar way, \cite{2018AJ....156..278G} measured 19 fields in both the LMC and the SMC to derive the mean TRGB position of $K=12.146$ and $(J-K_{\rm S})=1.147$ for the LMC, and $K=12.880$ and $(J-K)=1.021$ for the SMC. It can be seen that the values are slightly larger than ours. This may be caused by that the interstellar extinction is corrected for our sample stars. \cite{2021ApJ...923..232R} adopted the reddening value $E(V-I)$ from the 2D extinction map using the red clump stars by \cite{2021ApJS..252...23S}. On average, the foreground and internal extinction in the visual band $A_{V}$ is about 0.3-0.6 mag which means about 0.03-0.06 mag in $A_{K}$ and 0.05-0.09 in $E(J-K)$. Considering the difficulty in constructing the extinction map, the \cite{2022MNRAS.511.1317C} extinction map of the LMC is incorporated to examine the error brought about by the extinction correction. With this new extinction correction, the TRGB position becomes $K_{0}=12.000$  and $(J-K)_{0} = 0.991$, which indicates the error of extinction can lead to an uncertainty of about 0.01mag in $K_{\rm 0}$ and $(J-K)_{\rm 0}$, as expected from the insensitivity of near-infrared band to extinction.   Since the uncertainty of the extinction is small, the later analysis still uses the data with the extinction correction carried out by \cite{2021ApJ...923..232R}.

\subsection{Distribution of \texorpdfstring{$(J-K)_{0}$}{} and \texorpdfstring{$K_{0}$}{}}

As detailed in Section \ref{sec:Voronoi Binning}, the LMC and the SMC are divided into 154 and 70 bins respectively with about 300 and 120 stars within each bin by the Voronoi method. The values of $(J-K)_{0}$ and $K_{0}$ of each bin's TRGB are calculated by the PN method and their uncertainties by the bootstrap re-sampling. 
In the MCs, numerous star clusters are present. To assess their potential impact on determining the TRGB position, we scrutinized specific star clusters identified as "C" type in \cite{2022A&A...666A..80N}, \cite{1999AJ....117..238B}, and \cite{1995ApJS..101...41B}. The biggest cluster is NGC1866 with a size of 5.5 arcmin, which has only 9 member stars after cross-matching with our SampleL. In comparison with the about 300 objects in one bin in our analysis, these cluster stars have no impact on our determination of the TRGB.  In the SMC, NGC121 and K31 are cross-matched with SampleS, respectively. There are also very few member stars within the clusters or even none. The size of each pixel in the LMC is 0.3\degr, and 0.15\degr in the SMC, and each bin contains more than one pixel. However, the size of star clusters in their catalogue with type "C" is only a few arcmin. Moreover, many of them are open clusters that consist of blue stars, while our determination of TRGB relies on luminous red stars.

\begin{figure}
	\centering
    \includegraphics[scale=0.5]{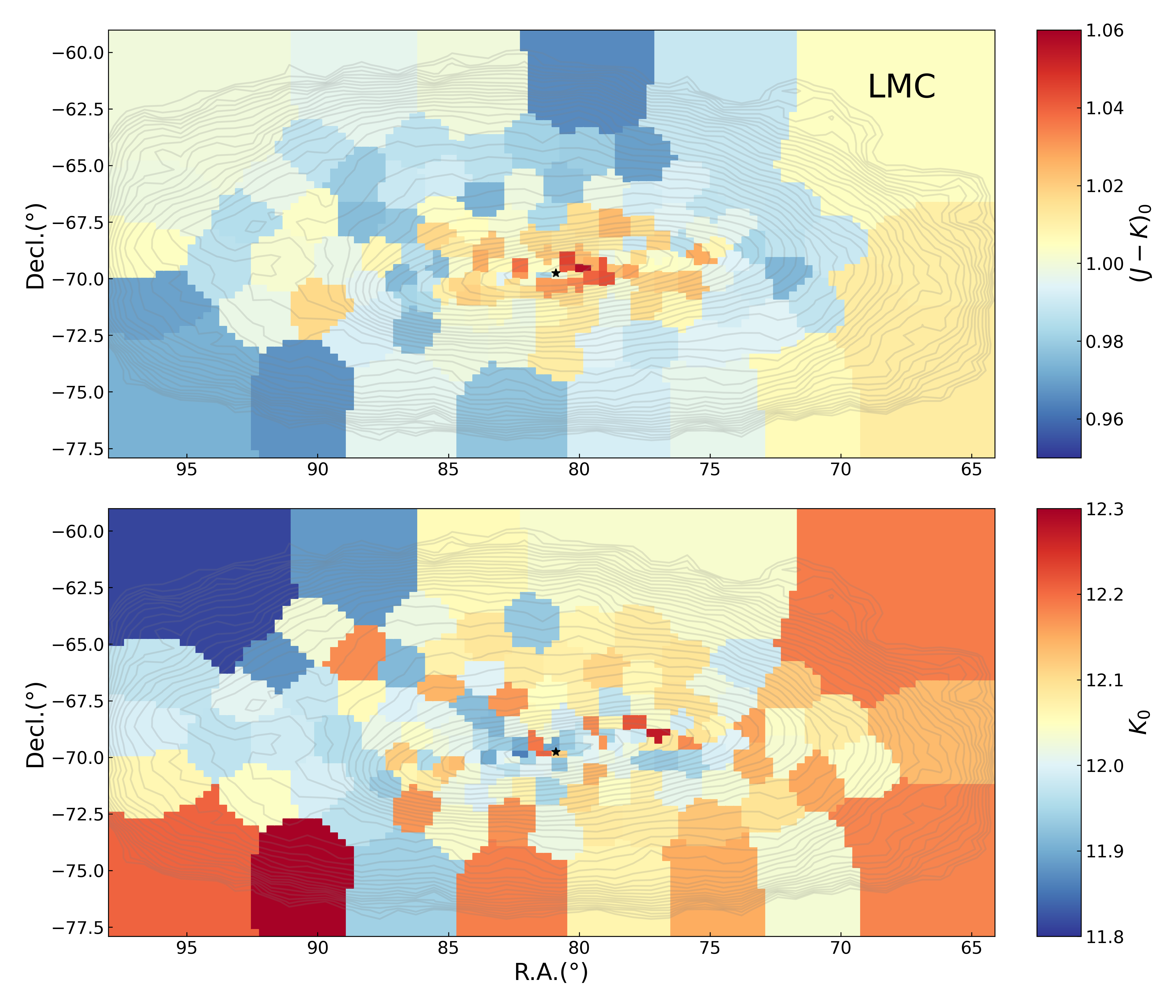}
	\caption{The $(J-K)_{0}$ and $K_{0}$ distribution in the LMC. The contour denotes the density of the member stars. The black asterisk marks the center of the LMC \citep{1976RC2...C......0D}. \label{fig: LMC-(J0-K0)-K0}}
\end{figure}

Figure \ref{fig: LMC-(J0-K0)-K0} shows the distribution of $(J-K)_{0}$ and $K_{0}$ in the LMC. It is clear that $(J-K)_{0}$ presents a gradually decreasing trend from the center to the periphery, and the center is bar-like. On the other hand, this trend is not so clear in the distribution of $K_{0}$. In the distribution of $K_{0}$, the increasing trend from the center to the east is weaker than to the west. In addition to metallicity and extinction, the brightness of TRGB, $K_{0}$, may be additionally influenced by the vertical depth in comparison with the color index $(J-K)_{0}$, while this work assumes a constant distance of 50 kpc for the LMC. Therefore, only the color index is used to infer the metallicity distribution, and the situation is the same for the SMC.

\begin{figure}
	\centering
    \includegraphics[scale=0.4]{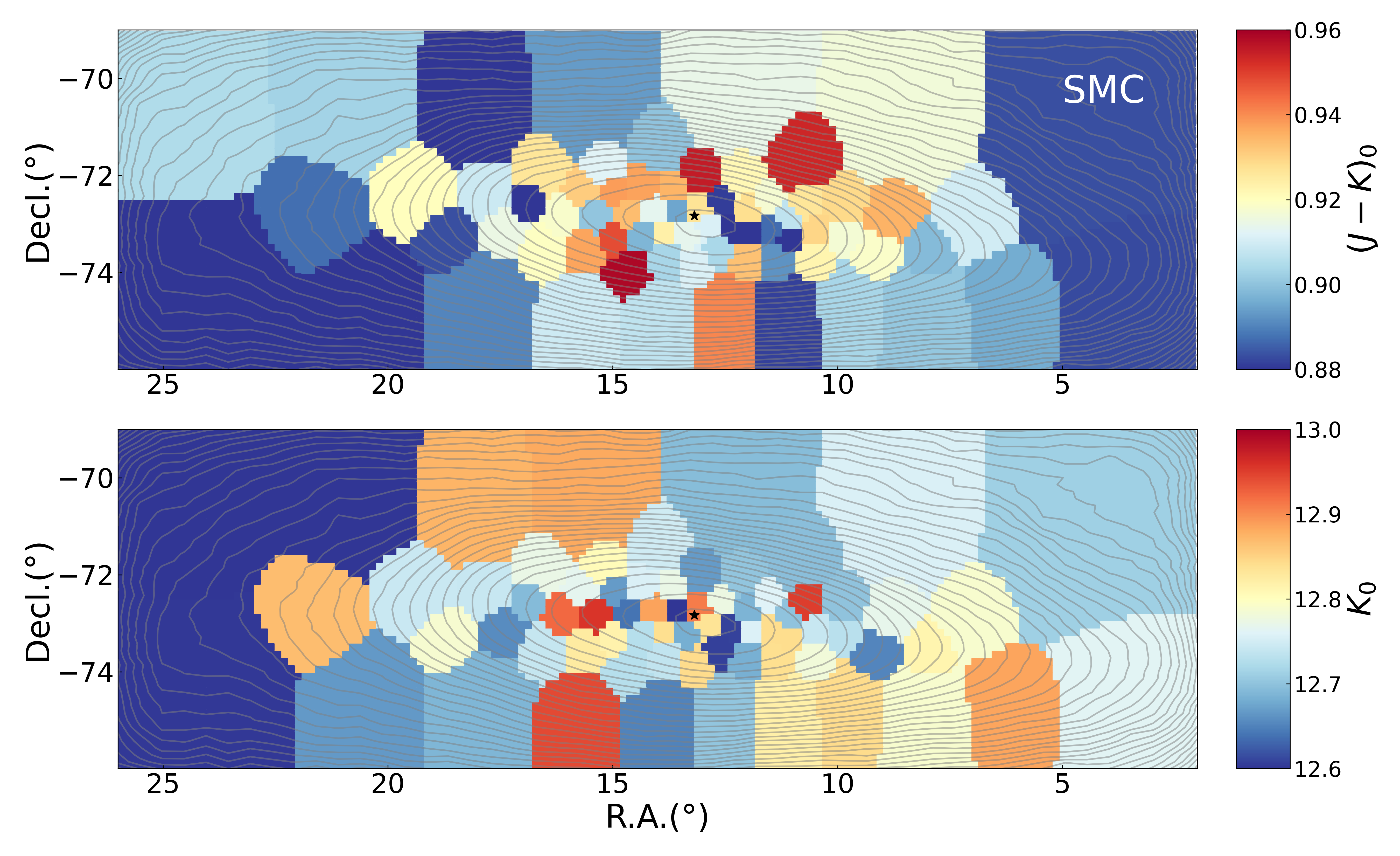}
	\caption{The same as Figure \ref{fig: LMC-(J0-K0)-K0}, but for the SMC. \label{fig: SMC-(J0-K0)-K0}}
\end{figure}

Figure \ref{fig: SMC-(J0-K0)-K0} shows the $(J-K)_{0}$ and $K_{0}$ distribution in the SMC.  In the distribution of $(J-K)_{0}$, some bins have higher $(J-K)_{0}$ near R.A. $\approx$ 15\degr and they form a semi-ring, it could be a star-forming region.

\subsection{Metallicity gradient}

The color index $(J-K)_{0}$ of TRGB is a metallicity indicator that increases with metallicity. As \cite{2000AJ....119..727B} point out the advantage of $J-K$ as a metallicity indicator is its insensitivity to reddening. There are a few works on the relation of the TRGB color and brightness with metallicity. \cite{2004A&A...424..199B} performed new empirical calibrations of the TRGB brightness in the $I$, $J$, $H$, and $K$ bands with the metallicity from the globular clusters, $\omega$ Centauri and 47 Tucanae, using a large optical and near-infrared photometric database. They yielded the relations: $M_{\rm I}^{\rm TRGB} = 0.258[M/H]^2 + 0.676[M/H] - 3.629$, $M_{\rm K}^{\rm TRGB} = -0.64[M/H] - 6.93$, $M_{\rm H}^{\rm TRGB} = -0.54[M/H] - 6.74$, and $M_{\rm J}^{\rm TRGB} = -0.26[M/H] - 5.55$. Meanwhile, \cite{2004MNRAS.354..815V} used the data from 24 Galactic globular clusters with homogeneous near-infrared photometry. They studied the luminosity of the RGB bump and tip in the $J$, $H$, and $K$ bands and their dependence on metallicity.
Comparison of the functional forms in these two studies finds that the relation of $(J-K)_{0}$ with metallicity agrees with each other. Consequently, we opt to use the relationship between $(J-K)_{0}$ and $[M/H]$ as presented by \cite{2004MNRAS.354..815V} to calculate the metallicity from $(J-K)_{0}$, i.e. $(J-K)_{0}=0.3*[M/H]+1.28$, for each bin.

\begin{figure}
	\centering
    \includegraphics[scale=0.7]{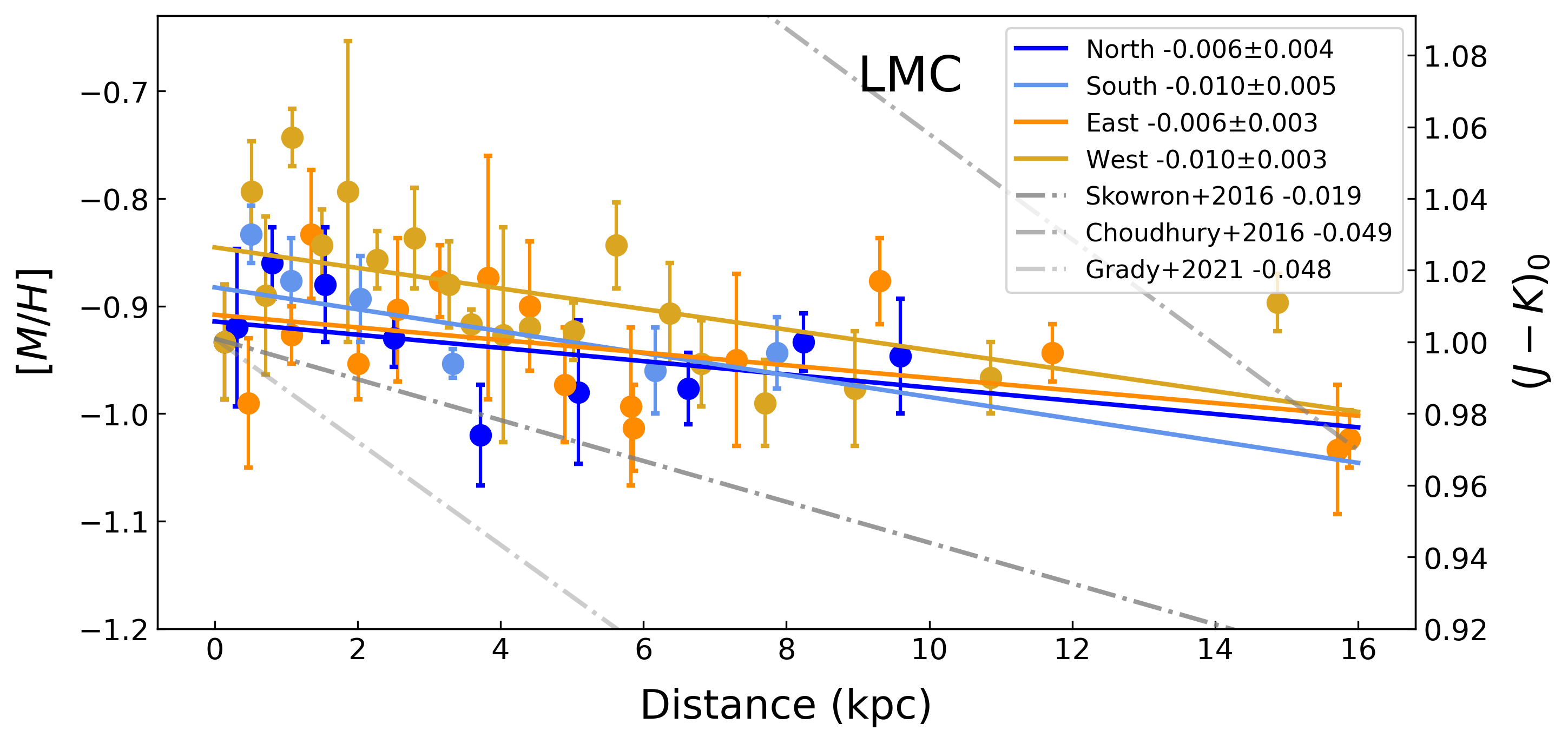}
	\caption{The variation of the metallicity with the distance from the galactic center in the LMC. The color codes the four directions. The slope of linear fitting for [M/H] and distance is shown in the upper right corner, which is compared with the literatures.   \label{fig: LMC-Z-dis}}
\end{figure}

\begin{figure}
	\centering
    \includegraphics[scale=0.7]{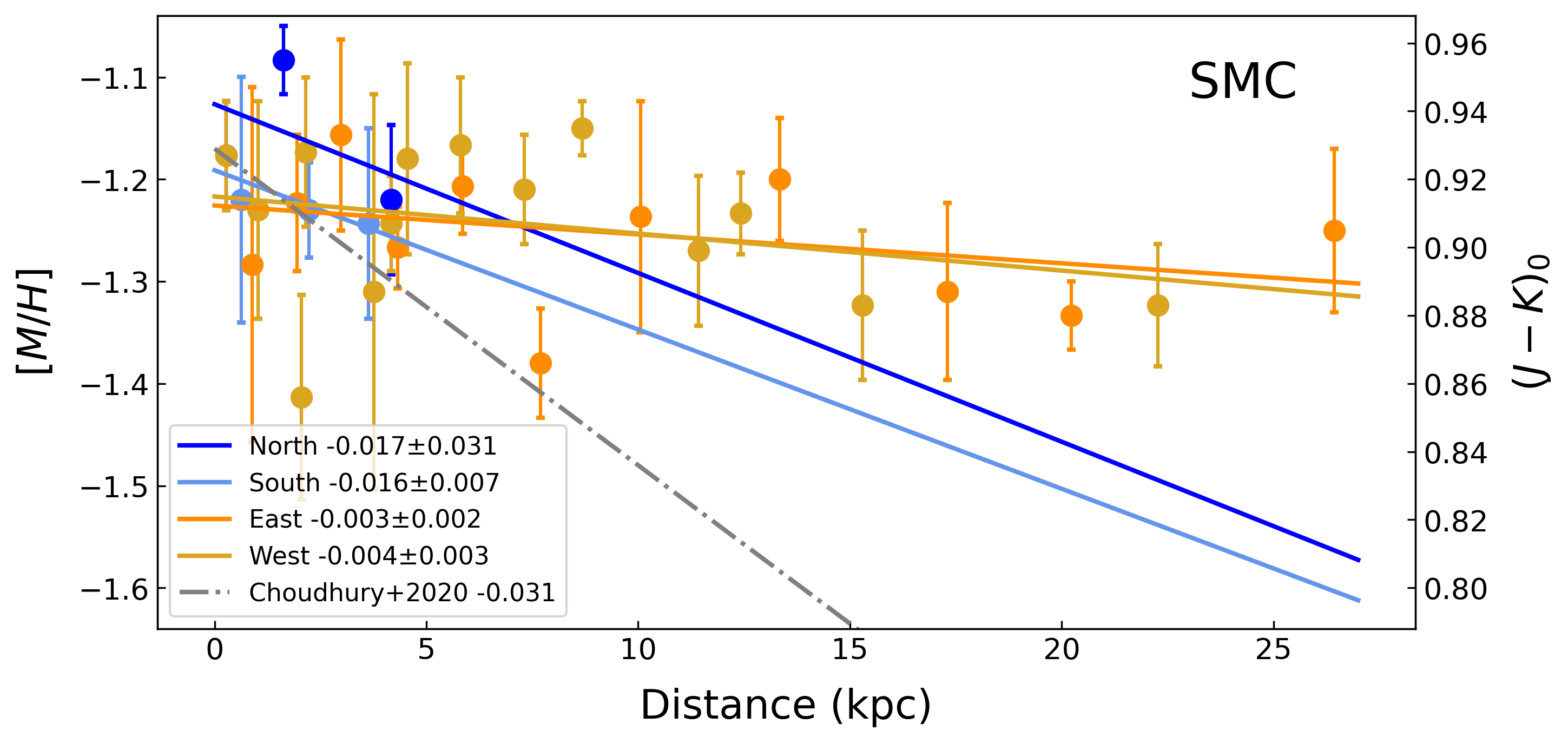}
	\caption{The same as Figure \ref{fig: LMC-Z-dis}, but for the SMC. \label{fig: SMC-Z-dis}}
\end{figure}

In order to illustrate the global features of the metallicity distribution, four typical directions are defined. For the LMC, the center is firstly located at $(\alpha, \delta)=(80.894\degr, -69.756\degr)$ \citep{1976RC2...C......0D}, and the line along the bar through the center is determined to be the east-west (EW) axis. Then, the line perpendicular to the EW axis is the north-south (NS) axis, which results in the position angle of $134.97\degr$ \citep{1989spce.book.....L}. For the SMC,  the center is at $(\alpha, \delta) = (13.187\degr, -72.829\degr)$ \citep{1976RC2...C......0D}, and the EW axis is aligned with the density contour distribution and the NS axis is perpendicular to the EW axis.
Figure \ref{fig: LMC-Z-dis} and Figure \ref{fig: SMC-Z-dis} display the variation of the metallicity with the distance from the center, where the adopted distance to the LMC and the SMC is 50 kpc and 61 kpc respectively \citep{2021MNRAS.507.4752C}. Because \citet{2021ApJ...909..150G} and \citet{2016AcA....66..269S} only provide the slope, the zero point is set to coincide with ours at the innermost distance. 
Following \cite{2009A&A...506.1137C}, the spherical coordinate and de-projection are considered to calculate the distance of each bin to the galaxy center. 
The inclination angle of the LMC is taken as 34.7\degr \citep{2001AJ....122.1807V}, and the position and inclination angle of the SMC is 45\degr and 65.5\degr respectively \citep{2009A&A...506.1137C}.
Overall, there is a negative metallicity gradient in each direction of the LMC and the SMC and the metallicity decreases with the distance from the center. For the LMC, the distance along the bar extends up to $\sim$16 kpc, and the north and south only extend up to $\sim$10 kpc. The least squares method is used for linear fitting, which results in a metallicity gradient to the north of $-0.006 \pm 0.004$ dex kpc$^{-1}$, to the south of $-0.010 \pm 0.005$ dex kpc$^{-1}$, to the east of $-0.006 \pm 0.003$ dex kpc$^{-1}$, and the west of $-0.010 \pm 0.003$ dex kpc$^{-1}$. This gradient, around -0.01 dex kpc$^{-1}$, is smaller than $\sim -0.05$ dex kpc$^{-1}$ by \citet{2009A&A...506.1137C}, \citet{2016MNRAS.455.1855C}, and \citet{2021ApJ...909..150G}, though shares the same decreasing tendency with the distance from the center. In addition, there is an obvious offset to the result of \citet{2016MNRAS.455.1855C}. For the SMC, the distance along the bar reaches up to roughly $\sim$27 kpc, while the north and south directions extend only up to about $\sim$4 kpc. The metallicity gradient is $-0.017 \pm 0.031$ dex kpc$^{-1}$ to the north, $-0.016 \pm 0.007$ dex kpc$^{-1}$ to the south, $-0.003 \pm 0.002$ dex kpc$^{-1}$ to the east and $-0.004 \pm 0.003$ dex kpc$^{-1}$ to the west. This result is similar to the value of $-0.031$ dex kpc$^{-1}$ by \citet{2020MNRAS.497.3746C}. Different from the case of the LMC, there is no systematic offset to the results of \citet{2020MNRAS.497.3746C}.

There is an ascending trend observed from the center to 1 kpc in the north, south, and west directions in the LMC, while a contrasting decreasing trend from the center to 1 kpc is noted in the south, east, and west directions in the SMC. It is essential to acknowledge that the slopes of pre-existing abundance gradients, encompassing both gas and stars, may undergo modifications due to the presence of a bar.
As \cite{1995A&A...301..649F} said, the stronger the bar, the weaker the gradient. But in the intense central star formation region, the gradient may steepen.

\section{Summary}

In order to delineate the metallicity structure of the Magellanic Clouds, this study takes advantage of the most pure and complete sample of red giant stars  which is achieved by excluding the foreground stars with the Gaia astrometric measurements as well as the near-infrared colors by \cite{2021ApJ...923..232R}. In addition, the color of TRGB is an accurate indicator of metallicity due to its insensitivity to interstellar reddening. Considering the very uneven distribution of stars, the Voronoi method is used to divide the area into the bins each of which has more or less equal number of stars, which leads to 154 bins with about 300 stars in each for the LMC and 70 bins with about 120 stars in each for the SMC. The position of TRGB in the $(J-K)_0/K_0$ diagram is detected by the Poisson noise weighted method, and the Bootstrap re-sampling is used to calculate the uncertainty.

For the LMC, the resultant TRGB is $K^{\rm TRGB}_{0}$ = $12.007 \pm 0.006$, $(J-K)^{\rm TRGB}_{0}$ = $1.002 \pm 0.001$, while for the SMC, it is $K^{\rm TRGB}_{0}$ = $12.707 \pm 0.006$, $(J-K)^{\rm TRGB}_{0}$ = $0.912 \pm 0.002$, which agree with previous results within about 0.01mag.
The gradient in the LMC of $(J-K)_{0}$ in the North, South, East, and West is $-0.0018 \pm 0.0013$ mag kpc$^{-1}$, $-0.0031 \pm 0.0016$ mag kpc$^{-1}$, $-0.0018 \pm 0.0008$ mag kpc$^{-1}$, and $-0.0029 \pm 0.0010$ mag kpc$^{-1}$, respectively. The gradient in the SMC of $(J-K)_{0}$ in the North, South, East, and West is $-0.0050 \pm 0.0093$ mag kpc$^{-1}$, $-0.0047 \pm 0.0022$ mag kpc$^{-1}$, $-0.0008 \pm 0.0007$ mag kpc$^{-1}$, and $-0.0011 \pm 0.0010$ mag kpc$^{-1}$, respectively.  Converting the color index of TRGB into metallicity, the metallicity gradients of the LMC and the SMC are obtained in four typical directions. For the LMC, the gradient to the north is $-0.006 \pm 0.004$ dex kpc$^{-1}$, to the south  $-0.010 \pm 0.005$ dex kpc$^{-1}$, to the east  $-0.006 \pm 0.003$ dex kpc$^{-1}$, and to the west  $-0.010 \pm 0.003$ dex kpc$^{-1}$. The farthest distance extends to 16 kpc. Our results agree well with the value of $-0.008 \pm 0.001$ dex kpc$^{-1}$ by \cite{2021MNRAS.507.4752C}. For the SMC, the gradients to the north, south, east, and west are $-0.017 \pm 0.031$ dex kpc$^{-1}$, $-0.016 \pm 0.007$ dex kpc$^{-1}$, $-0.003 \pm 0.002$ dex kpc$^{-1}$, and $-0.004 \pm 0.003$ dex kpc$^{-1}$, respectively. The farthest distance for the SMC extends to 27 kpc. The gradient is large from the center to 1 kpc, which may be related to the bar structure.

\acknowledgments{We thank Drs. Jian Gao, Haibo Yuan, Jun Li, and Zhetai Cao for their very helpful discussion. Thanks to Taylor Hoyt for the help in the Voronoi binning technique. We are also grateful to the anonymous referee for his/her useful comments.
This work is supported by the NSFC projects 12133002 and 12203016, National Key R\&D Program of China No. 2019YFA0405503, CMS-CSST-2021-A09.}

\vspace{5mm}
\facilities{}

\software{Astropy \citep{2013A&A...558A..33A}, TOPCAT \citep{2005ASPC..347...29T}
          }

\bibliography{paper}{}
\bibliographystyle{aasjournal}

\end{CJK*}
\end{document}